\documentclass{article}
\begin{document}
\begin{center}
{\Large Casimir Energy due to a Semi-Infinite Plane Boundary}
\end{center}
\vspace{5mm}
\begin{center}
H. Ahmedov*\footnote{E-mail: hagi@gursey.gov.tr} and I. H.
Duru**\footnote{E-mail: ihduru@galois.iyte.edu.tr}
\begin{flushleft}
   $\ast$) Feza Gursey Institute, P.O. Box 6, 81220, \c{C}engelk\"{o}y,
Istanbul, Turkey.
\\
 $\ast\ast$) Izmir Institute of Technology, Gulbahce 35437,
Urla-Izmir, Turkey.
\end{flushleft}

\end{center}
\begin{center}
{\bf Abstract}
\end{center}
Following the derivation of  the Green function for the massless
scalar field satisfying the Dirichlet boundary condition on the
Plane ( $x\geq 0$, $y=0$ ), we calculate the Casimir energy.

\vspace{1cm} \noindent {\bf 1. Introduction}
\\
\\
Thinking the open space Casimir energy calculations with  plane
boundaries we can name three essential examples: Single infinite
plane \cite{1}, the original parallel planes configuration \cite{2},
and the wedge problem, that is the two inclined planes geometry
\cite{3}. In this note we study a single semi-infinite plane
boundary.
\\
\\
In the following section we derive the exact propagator for massless
scalar field with Dirichlet boundary condition on the semi-infinite
plane.
\\
\\
Section III is devoted to the calculation of the Casimir energy.

\vspace{1cm} \noindent {\bf II. Green Function}
\\
\\
We need the Green function for massless scalar field satisfying
Dirichlet boundary condition on the semi-infinite plane given by
\begin{equation}
  P: \ { x\geq 0, \ \ y=0 }.
\end{equation}
Since the time and coordinate section of the massless Klein-Gordon
field is trivial we can write the Green function as ( with x
standing for the four dimensional  vector $x_\mu$ )
\begin{equation}
G(x,x^\prime )= \frac{1}{(2\pi)^2}\int dp_0dp_z
e^{-ip_0(t-t^\prime)}e^{-ip_z(z-z^\prime)}G_p(\vec{x},\vec{x}^\prime
)
\end{equation}
\\
Here $\vec{x}=(x,y)$ and $p=\sqrt{p_0^2+p_z^2}$. $G_p$ satisfies
\\
\begin{equation}\label{KG}
    (p^2 - \partial^2_x-\partial_y^2)G_p(\vec{x},\vec{x}^\prime
)=\delta(x-x^\prime)\delta(y-y^\prime)
\end{equation}
Note that we used Euclidean metric instead of Minkowski one. This is
always possible if the propagation of the fields in the direction of
time is free.  We now employ a variable change
\begin{equation}\label{tr}
    x=u^2-v^2, \ \ \ y=2uv, \ \ \ \ v\geq 0
\end{equation}
which maps the  plane $P$ onto
\begin{equation}
    \overline{P}: \  u=0.
\end{equation}
The full $(x,y)$ plane is mapped on the half plane $v\geq 0$.
\\
In  $u,v$ coordinates  (\ref{KG}) becomes
\begin{equation}
    (4p^2(u^2+v^2)  - \partial^2_u-\partial_v^2)\overline{G}_p(\vec{u},\vec{u}^\prime
)=\delta(u-u^\prime)\delta(v-v^\prime)
\end{equation}
which is the equation for two noninteracting harmonic oscillators,
with $v\geq 0$ and Dirichlete boundary condition at $v=0$. The Green
function $\overline{G}_p(\vec{u},\vec{u}^\prime )$ corresponds to
the Kernel satisfying
\begin{equation}
    (\frac{d}{dS}+ (-\frac{1}{2}\partial^2_u + \omega^2u^2 ) + (-\frac{1}{2}\partial^2_v + \omega^2v^2 ) )K(\vec{u},\vec{u}^\prime
)=\delta(S)\delta(u-u^\prime)\delta(v-v^\prime)
\end{equation}
where $S$ is the Euclidean "time" interval $s-s^\prime$ and
$\omega\equiv 2p$. The Euclidean "time" oscillator Kernel is given
by the well known formula \cite{4}:
\begin{equation}\label{Ker}
    K(\vec{u},\vec{u}^\prime ;S) =\Theta (S)K_1(u,u^\prime ; S)K_2(v,v^\prime ; S)
\end{equation}
where
\begin{equation}
   K_1(u,u^\prime ; S)=\sqrt{\frac{\omega}{2\pi\sinh\omega S}}
   Exp [-\frac{\omega}{2\sinh\omega S}((u^2+u^{\prime 2})\cosh\omega S-2uu^\prime
   )]
\end{equation}
\begin{eqnarray}\label{10}
% \nonumber to remove numbering (before each equation)
K_2(v,v^\prime ; S)=\sqrt{\frac{\omega}{2\pi\sinh\omega S}} (
   Exp [-\frac{\omega}{2\sinh\omega S}((v^2+v^{\prime 2})\cosh\omega S-2vv^\prime
   )] - \nonumber \\
- Exp [-\frac{\omega}{2\sinh\omega S}((v^2+v^{\prime 2})\cosh\omega
S +2vv^\prime
   )])
\end{eqnarray}
By construction the Kernel $K_2$ vanishes at $v=0$. The Green
function $\overline{G}_p(\vec{u},\vec{u}^\prime )$  is simply the
Fourier transform of the Kernel of (\ref{Ker})
\begin{equation}
    \overline{G}_p(\vec{u},\vec{u}^\prime
    )=\frac{1}{2}\int_0^\infty dS
    K(\vec{u},\vec{u}^\prime ; S).
\end{equation}
After dropping the first term in (\ref{10}) which corresponds to the
free Minkowski space term, we get the regularized Green function
\begin{equation}
    \overline{G}^R_p(\vec{u},\vec{u}^\prime
    )=-\frac{1}{2}\int_0^\infty dS \frac{\omega}{2\pi\sinh\omega S}
Exp[-\frac{\omega}{2\sinh\omega S}(\xi\cosh\omega S - \eta )]
\end{equation}
where
\begin{equation}
   \xi=u^2+u^{\prime 2}+v^2+v^{\prime 2}, \ \ \  \eta = 2(uu^\prime -vv^\prime
   ).
\end{equation}
After inserting the above expression into (2) in place of
$G_p(\vec{x},\vec{x}^\prime )$ we arrive at the regularized Green
function $G^R (x, x^\prime)$ which vanishes on the semi-infinite
plane P of (1).

\vspace{1cm} \noindent {\bf III. Casimir Energy density }
\\
\\
To obtain the Casimir energy density we employe the well known
coincidence limit formula \cite{5} which in the Euclidean metric
reads
\begin{equation}
   T(x)=\frac{1}{2}\lim_{x\rightarrow x^\prime} (-\partial_t\partial_{t^\prime}+ \partial_z\partial_{z^\prime}+
   \partial_x\partial_{x^\prime}+\partial_y\partial_{y^\prime})G^R(x,x^\prime )
\end{equation}
or in $u,v$ coordinates
\begin{equation}
   T(x)=\frac{1}{2}\lim_{x\rightarrow x^\prime} (-\partial_t\partial_{t^\prime}+ \partial_z\partial_{z^\prime}+
   \frac{1}{4(u^2+v^2)}(\partial_x\partial_{x^\prime}+\partial_y\partial_{y^\prime}))G^R(x,x^\prime )
\end{equation}
Since the system is invariant under the interchange of $t$ and $z$
after integration over $dp_0$ and $dp_z$ the  first two terms cancel
each other. This fact gives an opportunity to simplify our
expression further. Instead of $G^R(x,x^\prime)$ we can work with an
effective Green function
\begin{equation}
G_{eff}^R(\vec{x},\vec{x}^\prime)=G^R(x,x^\prime)\mid_{t=t^\prime,
z=z^\prime}
\end{equation}
which is equal to
\begin{equation}
   G_{eff}^R(\vec{x},\vec{x}^\prime)=\frac{1}{2\pi}\int_0^\infty pdp\overline{G}^R_p(\vec{u},\vec{u}^\prime)
\end{equation}
Inserting (12) in the above equation in place of
$\overline{G}^R_p(\vec{x},\vec{x}^\prime)$ we can first integrate
over $dp$ ($\omega$ in (12) is $\omega=2p$ ), then over $dS$. We
then obtain
\begin{equation}
   G_{eff}^R(\vec{x},\vec{x}^\prime)=-\frac{1}{8\pi^2 \xi} \frac{1}{(u-u^\prime)^2+(v+v^\prime)^2}
\end{equation}
Using this simple formula in (15) (with first two terms dropped) as
the Green function we get expression for the energy density
\begin{equation}
   T(x) = -\frac{1}{128\pi^2v^2(u^2+v^2)^2} (\frac{1}{2v^2}+\frac{1}{(u^2+v^2)})
\end{equation}
or in terms of the original $x,y$ coordinates ( with
$r=\sqrt{x^2+y^2}$ )
\begin{equation}
   T(x) = -\frac{1}{64\pi^2r^2(r-x)} (\frac{1}{r-x}+\frac{1}{r}),
\end{equation}
which in terms of polar coordinates $x=r\cos\theta, \ y=r\sin\theta$
is
\begin{equation}
   T(x) = -\frac{1}{64\pi^2r^4(1-\cos\theta)}
   (\frac{1}{(1-\cos\theta)}+1).
\end{equation}
Inspecting (20) we see that at $x>0$ and $y\ll x$ we obtain the
infinite plane result \cite{5}
\begin{equation}
T\simeq -\frac{1}{16\pi^2y^4}; \ \ \ x>0,  y\ll x
\end{equation}
as expected.

 \vspace{1cm}\noindent {\bf Acknowledgments} Authors thank
Turkish Academy of Sciences (TUBA) for its support and Dr. O.
Pashaev for discussion.

\end{document}